\documentstyle{npbproc}

\def\3{\ss}
\newcommand{\D}{\not{D}}
\newcommand{\DD}{\not\!\!{D}}
\newcommand{\p}{\not\!{\partial}}
\newcommand{\A}{\not\!\!{A}}
\newdimen\fmmargin \fmmargin=4em
\begin{document}

\vspace{-0.5cm}
\title{
       A PROPOSAL FOR SIMULATING CHIRAL FERMIONS\thanks{Talk given
at the {\em Workshop on Non-Perturbative Aspects of Chiral Gauge
Theories}, Rome, March 1992}}

\author{ M. G\"ockeler$^{1,2}$ and G. Schierholz$^{2,3}$
 \\[2em]
$^1$ Institut f\"ur Theoretische Physik, RWTH Aachen,\\[0.2em]
Sommerfeldstra\3e, D-5100 Aachen, Germany\\[0.6em]
$^2$Gruppe Theorie der Elementarteilchen,
H\"ochstleistungsrechenzentrum HLRZ, \\  [0.2em]
c/o KFA, Postfach 1913, D-5170 J\"ulich, Germany\\[0.6em]
$^3$ Deutsches Elektronen-Synchrotron DESY,\\[0.2em]
Notkestra\3e 85, D-2000 Hamburg 52, Germany}

\date{}

\runtitle{Chiral Fermions}
\runauthor{M. G\"ockeler and G. Schierholz}

%\pubyear
\volume{25A}
\firstpage{1}
\lastpage{10}
\newdimen\fmmargin \fmmargin=0em

\begin{abstract}  \vspace{1.0cm}
Abstract:
We describe a method for evaluating chiral gauge theories that is not
plagued by the doubling problem. To demonstrate the efficiency of the
approach, we apply our ideas
to the chiral Schwinger model.

\end{abstract}

\maketitle

\section{ INTRODUCTION }

Chiral gauge theories play an important role in particle physics.
Yet very little is known about their properties. The main reason
for this is the lack of a calculational method which takes into
account non-perturbative effects.

The lattice regularization was invented for that purpose. Until
now we have, however, not been able to find a
consistent lattice formulation of these theories. Due to the
doubling problem one is always led to an equal number of left- and
right-handed fermions with net chirality zero. Attempts to remove
the unwanted ``doublers'' by pushing their masses to the cut-off while
preserving chiral symmetry have, in spite of great efforts, not
brought the desired success \cite{Gol}.
Some authors have arranged themselves with
this situation by postulating the existence
of mirror fermions \cite{Mon}. But
even these models are inadequate. A non-vector theory
leads - after the anomalies have been cancelled - to a gauge invariant,
complex effective action whose imaginary part is
non-zero \cite{Alv}. The action of
the mirror-fermion model is, however, real.

But it may also be that non-vector theories do not exist as consistent
field theories with non-trivial dynamics. A first answer to this
question can be obtained from the chiral Schwinger model. In four
dimensions one has to resort to numerical methods to address this
problem. We believe, however, that it should at least be possible to
answer some of the more fundamental questions, in particular whether
the imaginary part of the effective action is indeed non-zero.

In order to evade the doubling problem, we propose to evaluate the
fermionic action in the background of continuum gauge fields, which we
will construct from lattice gauge field configurations.
The starting point is a formula for the effective action, which was
derived some time ago by Alvarez-Gaum\'{e} et al. \cite{Alv} for a
similar purpose. The virtue of this formula is that it involves the
ordinary covariant Dirac operator only and that it is amenable to
explicit calculations. The basic idea of the method as well as some
theoretical background are presented in sec. 2. In sec. 3 we sketch
the derivation of continuum gauge fields from lattice gauge field
configurations. Our ideas are then applied to the chiral Schwinger
model in sec. 4. Finally, in sec. 5 we present some concluding
remarks.

\section{EFFECTIVE ACTION}

We will work in Euclidean space compactified to the torus {\bf T}$^4$.
But one can also think of other geometries. Consider now the Dirac
operator
\begin{equation}
i\DD_{\mp}(A) = i(\p + \A P_{\mp}),\; P_{\mp} = \frac{1 \mp %
\gamma_5}{2},
\end{equation}
which minimally couples left-handed and right-handed fermions,
respectively, to the gauge fields. The gauge fields are assumed to be
in a complex representation of some gauge group $G$,
\begin{equation}
A_{\mu} = A_{\mu}^a T^a, A_{\mu}^\dagger = -A_{\mu}.
\end{equation}
We are interested in the effective action for left-handed fermions,
$S_{-}(A)$. This is given by
\begin{equation}  \label{eq23}
e^{-S_{-}(A)} = \int {\cal D}\bar{\psi} {\cal D}\psi e^{-\bar{\psi}%
i\D_{-}(A)\psi}.
\end{equation}
It is understood that eq. (\ref{eq23}) and the following expressions are
properly regularized, without committing ourselves to any particular
scheme for the moment. For the real part of $S_{-}(A)$ one finds
\begin{equation}
2 {\rm Re} S_{-}(A) = S_{-}(A) + S_{+}(A) = S(A) + S(0)
\end{equation}
with
\begin{equation}
e^{-S(A)} = \int {\cal D}\bar{\psi} {\cal D}\psi e^{-\bar{\psi}%
i\D(A)\psi},
\end{equation}
\begin{equation}
i\DD(A) = i(\p + \A),
\end{equation}
which means that it is given by the action of the corresponding vector
theory. This part of the action does not constitute any problems. It
can be computed by standard lattice techniques.

Thus the chiral nature of the fermions is reflected entirely in the
{\em imaginary} part of the effective action.
Following Alvarez-Gaum\'{e} et al. \cite{Alv}
we shall now derive an expression for Im$S_{-}(A)$ on which
we will base our further discussion. Our derivations will be purely
formal. For rigorous proofs the reader is refered to the original
literature \cite{Alv}. We start from the expression
\begin{equation} \begin{array}{l}
S_{-}(A)-S_{-}(0)  \\ [1.1em]
=  - {\rm ln\; det}\;i\DD_{-}(A)+{\rm ln\; det}\;i\DD_{-}(0)\\ [1.1em]
=  - {\rm Tr\; ln}\;i\DD_{-}(A) + {\rm Tr\; ln}\;i\DD_{-}(0)\\ [1.1em]
=  - {\rm Tr\; ln}\;(1+(i\p)^{-1}i\A P_{-})\; .
\end{array} \end{equation}
Expanding the logarithm, this gives
\begin{equation}  \label{eq28} \begin{array}{l}
S_{-}(A)-S_{-}(0) \\ [1.1em]  \displaystyle
= {\rm Tr} \sum_{n=1}^{\infty}\frac{(-1)^n}{n}%
((i\p)^{-1}i\A)^n P_{-}.
\end{array} \end{equation}
By replacing $A$ by $^{t}\!A = tA$ on the right-hand
side of eq. (\ref{eq28}),
differentiating the whole expression with respect to $t$ and
integrating over $t$ again, one obtains after doing the sum over $n$
again
\begin{equation} \label{eq29} \begin{array}{l} \displaystyle
S_{-}(A)-S_{-}(0) \\ [1.1em] \displaystyle
= - \int_0^1 dt \; {\rm Tr}\; i \A%
(i\DD(^{t}\!A))^{-1}P_{+}.
\end{array} \end{equation}
This results in the expression for the imaginary part
\begin{equation}  \label{eq210}
{\rm Im}S_{-}(A) = - \frac{1}{2} \int_0^1 dt \; {\rm Tr}\; \gamma_5 \A%
(i\DD(^{t}\!A))^{-1}.
\end{equation}

While the real part of the action is gauge invariant, the imaginary part
is generally not. Under gauge transformations
\begin{equation}
A_\mu \rightarrow ^{g}\!A_\mu = g^{-1}(A_\mu +\partial_\mu)g
\end{equation}
it transforms like
\begin{equation}
{\rm Im}S_{-}(^{g}\!A) - {\rm Im}S_{-}(A) = 2\pi Q_5,
\end{equation}
where $Q_5$ is the Wess-Zumino action.
If the theory is anomaly-free, $Q_5 = 0$ and
$S_{-}(A)$ is gauge invariant. But we still expect ${\rm Im}S_{-}(A)%
\neq 0$, also in a model with mirror fermions.

A straightforward lattice calculation of eq. (\ref{eq210}) with, e.g.,
Wilson fermions would give a divergent result in the continuum
limit.
The situation is similar to the case of the
topological charge,
\begin{equation} \label{eq213}
Q = - \frac{1}{16\pi^2} \int d^4x \;{\rm Tr}\; F \tilde{F},
\end{equation}
\begin{equation}
F \tilde{F} = \frac{1}{2} \epsilon_{\mu \nu \rho \sigma} F_{\mu \nu} %
F_{\rho \sigma},
\end{equation}
where the naive latice transcription of $F \tilde{F}$ \cite{DiV}
does not have any topological significance.
Indeed, it has been shown \cite{Alv} that the gauge
invariant part of ${\rm Im}S_{-}(A)$ is given by the $\eta$-invariant
of the five-dimensional Dirac operator,
\begin{equation}
H =  i\gamma_5\partial_t + i\DD(^{t}\!A),
\end{equation}
which is a topological invariant.

Let us stay with the example of the topological charge for a moment. It
is well known that the naive expression for $Q$ receives perturbative
contributions, which are divergent in the continuum limit and so cover
all topological effects. The origin of these contributions lies in
fluctuations of the gauge fields on the scale of the lattice spacing.
These fluctuations are unphysical near the continuum limit, and only
those theories which have a continuum limit are considered here.
If we now extrapolate the lattice gauge field
smoothly to the interior of the hypercubes - thereby respecting the
fact that a topologically non-trivial gauge field configuration cannot
be in the same gauge everywhere on the torus - and calculate $Q$ in
the background of this continuum gauge field using the expression
(\ref{eq213}),
we find not only that all divergences have
been eliminated\footnote{We assume that the action has been chosen
such that the susceptibility is not affected by
dislocations \cite{Goc}.},
but also that $Q$ is an integer as it should. (See
the next section for further details.) This is the essential idea
behind the geometrical expressions for the
topological charge \cite{Lue,Phi,Lau},
though the actual
calculation of $Q$ proceeds via the transition functions, which connect
the gauges in the different hypercubes, and for those it is sufficient
to know the gauge field on the {\em boundary}
of the hypercubes only. Given
the continuum gauge field, it would equally well be possible
to compute the
topological charge by discretizing this field again, but on a finer
mesh, computing the corresponding link matrices and using the
naive lattice expression for $F \tilde{F}$ \cite{DiV} to obtain $Q$.
The continuum value of the charge
is then found by making the mesh finer and finer until the result has
converged.

We now propose to compute ${\rm Im}S_{-}(A)$ in an analogous way. We
start from a lattice gauge field configuration, which we regard for the
moment as being given to us. The first
problem is then to extrapolate the lattice
gauge field to the {\em whole}
interior of the hypercube. This construction,
which is highly non-trivial, will be sketched in the next section.
Further details can be found in the original publication \cite{Sch}.
The next problem
is to compute ${\rm Im}S_{-}(A)$ in the background of this field.
As we have already indicated, the expressions for the effective action
have to be properly regularized in order to make sense.
We suggest to use the lattice regularization with a lattice
spacing, $a$,
much smaller than the original lattice spacing. The latter we shall set
equal to one for convenience.
For Wilson fermions we then find
\begin{equation}  \label{eq214} \begin{array}{l} \displaystyle
{\rm Im}S_{-}(A) =  \frac{i}{2} \int_{0}^1 dt \; a^4
\sum_{x,\mu} \{{\rm Tr}\;%
\gamma_5(\gamma_\mu - 1) \\ [1.1em] \displaystyle
\times \frac{1}{2a} {\rm ln}\;{\cal U}_{x,\mu}\;{\cal U}_{x,\mu}^t%
{\cal G}_{x+\hat{\mu},x} + {\rm Tr}\; \gamma_5(\gamma_\mu + 1) \\
[1.1em]  \displaystyle
\times \frac{1}{2a} {\rm ln}\;{\cal U}_{x,\mu}\;{\cal U}_{x,\mu}^{-t}%
{\cal G}_{x,x+\hat{\mu}}\},
\end{array} \end{equation}
where ${\cal U}$ are the link matrices on the fine
sublattice, and ${\cal G}$ is
the inverse fermion matrix,
\begin{equation} \begin{array}{l} \displaystyle
\frac{1}{2a}\sum_{\mu}
(\gamma_\mu - 1){\cal U}_{x,\mu}^t {\cal G}_{x+\hat{%
\mu},y} \\ [1.1em]  \displaystyle
- \frac{1}{2a}\sum_{\mu}(\gamma_\mu + 1) {\cal U}_{x-\hat{\mu}%
,\mu}^{-t} {\cal G}_{x-\hat{\mu},y} \\ [1.1em]  \displaystyle
+ \frac{4}{a} {\cal G}_{x,y} \} = \frac{1}{a^4} \delta_{x,y}\;{\bf 1}.
\end{array} \end{equation}
It is straightforward to compute
the link matrices ${\cal U}$, as well as the
logarithm and powers of ${\cal U}$, from the continuum gauge field.
The right-hand side of eq. (\ref{eq214}) can be very
efficiently calculated by means of a stochastic estimator \cite{Bit}.
This requires only one inversion of the fermion matrix for one set of
random numbers.
In the final step we then have to
perform the calculations on finer and finer
sublattices until we obtain a stable result (if this exists).

We like to emphasize that eq. (\ref{eq214}) is only one out of many
possibilities of computing ${\rm Im}\;S_{-}(A)$ in the background
of a continuum field. An alternative method would be to use the gauge
field $A_{\mu}(x)$ directly
instead of ${\rm ln}\;{\cal U}_{x,\mu}\;{\cal
U}_{x,\mu}^t$. But this would require a further regularization step.
One such step is the point-splitting technique.
Which method is best suited can
only be decided in concrete calculations which we will return to in
the future.

It may be possible that there are more elegant ways of computing the
imaginary part of the effective action. Alvarez-Gaum\'{e}
has attempted a geometric construction similar to that of the
topological
charge, but without success \cite{Pri}.
In two dimensions it turns out that one
can calculate the effective action analytically, as we will show in
sec. 4.

Whether we can do simulations of chiral gauge theories depends now on
the magnitude of ${\rm Im}S_{-}(A)$. So far we do not know  how to
handle complex actions. If ${\rm Im}S_{-}(A)$ is small, we may simulate
gauge field configurations according to the real part of the effective
action (including the gauge field action) on the lattice in the
standard fashion and include the phase factor in the statistical average
over the ensemble.

\section{CONTINUUM GAUGE FIELDS}

We shall now sketch the construction of continuum gauge fields
$A_\mu(x)$
from lattice gauge field configurations. We work on a hypercubic
lattice with periodic boundary conditions. The lattice points are
denoted by $s$, and the link matrices are called $U_{s,\mu}$.

The gauge fields have to fulfill the following constraints. (i) The
link matrices derived from $A_\mu$ must agree with those of the
original lattice, $U_{s,\mu}$. (ii) The construction must be gauge
covariant, which means that a lattice gauge field transformation must
result in a continuum gauge field transformation of $A_{\mu}(x)$.
(iii) The gauge field $A_{\mu}(x)$ should lead to the fiber bundle
which has been previously derived from the lattice gauge field
configuration \cite{Lue}.
This guarantees that the topological charge is the same
in both cases.

In the first step we construct in each hypercube,
\begin{equation}
c(s) = \{ x \in {\bf T}^4 \mid s_\mu \leq x_\mu \leq s_\mu + 1,
\forall \mu \},
\end{equation}
a local gauge field $A_{\mu}^{(s)}(x)$,
such that on the intersection of
two adjacent hypercubes the corresponding gauge fields are connected
by a gauge transformation given by L\"uscher's transition
functions \cite{Lue}.
In the second step we then construct a global, and generally singular,
gauge field out of the local gauge fields.

Because the construction involves derivatives of the transition
functions in intermediate steps, it is necessary to enlarge the
hypercubes to open sets,
\begin{equation}
\tilde{c}(s) = \{ x \in {\bf T}^4 \mid s_\mu -\epsilon <%
x_\mu < s_\mu + 1 + \epsilon,
\forall \mu \},
\end{equation}
and define smeared transition functions on open faces
$\tilde{f}(s,\mu) = \tilde{c}(s)\cap \tilde{c}(s-\hat{\mu})$:
\begin{equation} \begin{array}{l}
\tilde{v}_{s,\mu}(x) = v_{s,\mu}(s_1 +\phi(x_1 -s_1),\cdots \\
[1.1em]
\cdots,  s_\mu,\cdots,%
s_4 +\phi(x_4 -s_4)).
\end{array} \end{equation}
Here $v_{s,\mu}$ is L\"uscher's transition
function \cite{Lue} defined on
$f(s,\mu) = c(s)\cap c(s-\hat{\mu})$ and $\phi$ is a smooth function,
$\phi: {\bf R} \rightarrow [0,1]$, with the properties
\begin{equation}
\phi(t) = \left\{ \begin{array}{ll}
             0 & \mbox{for $t<\eta$}\\
             1 & \mbox{for $t>1-\eta$}
             \\
             \end{array}
             \right.
\end{equation}
and $\epsilon < \eta < 1/2$. It follows that the $\tilde{v}$'s satisfy
the cocycle condition
\begin{equation}
\tilde{v}_{s-\hat{\mu},\nu}(x) \tilde{v}_{s,\mu}(x) =
\tilde{v}_{s-\hat{\nu},\mu}(x) \tilde{v}_{s,\nu}(x)
\end{equation}
on $\tilde{p}(s,\mu,\nu)$ = $\tilde{f}(s,\mu)\cap \tilde{f}(s-\hat{\nu}%
,\mu)$ and therefore define a principal bundle over ${\bf T}^4$. The
gauge field $A_{\mu}^{(s)}(x)$ can now be constructed step by step
making use of the relation
\begin{equation}  \label{eq36} \begin{array}{l}
A_\mu^{(s-\hat{\nu})}(x) = \tilde{v}_{s,\nu}(x) A_\mu^{(s)}(x)
\tilde{v}^{-1}_{s,\nu}(x) \\ [1.1em]
+ \tilde{v}_{s,\nu}(x) \partial_\mu
\tilde{v}^{-1}_{s,\nu}(x).
\end{array} \end{equation}
Starting from the value zero at $x\in \tilde{c}(s)$, $-\epsilon < x_\nu%
 -s_\nu < \epsilon$ for $\forall \nu$,
eq. (\ref{eq36}) gives the gauge field
in the neighborhoods of the other corners of $\tilde{c}(s)$, where it is
also zero. In the next step one takes the gauge field to be zero in
$\epsilon$-neighborhoods of the links in $\tilde{c}(s)$ which
originate from $s$ and applies
eq. (\ref{eq36}) again.
Then one defines $A_{\mu}^{(s)}(x)$ on the six plaquettes
$\tilde{p}(s,\mu,\nu)$ by interpolating the expressions already
obtained. By means of eq. (\ref{eq36})
one finds $A_{\mu}^{(s)}(x)$ on all the
other plaquettes contained in $\tilde{c}(s)$. Continuing in this fashion
one arrives at the expression
%\begin{equation} \label{eq37}  \displaystyle
%\begin{equation} \label{eq37}  \begin{array}{l}
$$
\begin{array}{l} \displaystyle
A_\mu^{(s)}(x) = \phi(x_\alpha - s_\alpha) \phi(x_\beta -s_\beta)
\phi(x_\gamma - s_\gamma) \\ [1.1em]  \displaystyle
\times     \{ Z_\mu(s,\alpha,\beta,\gamma \mid s_\alpha + 1,
s_\beta + 1,s_\gamma + 1,x_\mu)
\end{array}
$$
$$
\begin{array}{l}  \displaystyle
+ \sum_{\stackrel{\rm cycl.\, perm.}{(\alpha,\beta,
\gamma)}} [ -{\rm Ad} ((\tilde{v}_{s+\hat{\alpha},\alpha}\\ [1.1em]
\displaystyle \times
\tilde{v}_{s+\hat{\alpha}+\hat{\beta},\beta})(s_\alpha + 1,s_\beta
+ 1,x_\gamma,x_\mu) )
\\ [1.1em]  \displaystyle
\times Z_\mu(s+\hat{\alpha}+\hat{\beta},\gamma \mid
s_\alpha + 1,s_\beta + 1,s_\gamma + 1,x_\mu)
\end{array}
$$
$$
\begin{array}{l} \displaystyle
+{\rm Ad} (\tilde{v}_{s+\hat{\alpha},\alpha}(s_\alpha + 1,
x_\beta, x_\gamma, x_\mu)
\\ [1.1em]  \displaystyle  \times
\tilde{v}_{s+\hat{\alpha}+\hat{\beta},\beta}(s_\alpha + 1,s_\beta
+ 1,x_\gamma,x_\mu) )
\\ [1.1em]  \displaystyle    \times
Z_\mu(s+\hat{\alpha}+\hat{\beta},\gamma \mid
s_\alpha + 1,s_\beta + 1,s_\gamma + 1,x_\mu)
\end{array}
$$
$$
\begin{array}{l} \displaystyle
+{\rm Ad} (\tilde{v}_{s+\hat{\alpha},\alpha}(s_\alpha + 1,
x_\beta, x_\gamma, x_\mu)
\\ [1.1em]  \displaystyle  \times
\tilde{v}_{s+\hat{\alpha}+\hat{\gamma},\gamma}(s_\alpha + 1,x_\beta,
s_\gamma + 1,x_\mu) )
\\ [1.1em]  \displaystyle
\times Z_\mu(s+\hat{\alpha}+\hat{\gamma},\beta \mid
s_\alpha + 1,s_\beta + 1,s_\gamma + 1,x_\mu)
\end{array}
$$
$$
\begin{array}{l} \displaystyle
-{\rm Ad} (\tilde{v}_{s+\hat{\alpha},\alpha}(s_\alpha + 1,
x_\beta, x_\gamma, x_\mu) )
\\ [1.1em]  \displaystyle
\times Z_\mu(s+\hat{\alpha},\beta,\gamma \mid
s_\alpha + 1,s_\beta + 1,s_\gamma + 1,x_\mu) ]       \}
\end{array}
$$
\begin{equation} \label{eq37}  \begin{array}{l}
\displaystyle
 + \sum_{\stackrel{\rm cycl.\, perm.}{(\alpha,\beta,\gamma)}}
[ - \phi(x_\alpha - s_\alpha) \phi(x_\beta -s_\beta)
\\ [1.1em]  \displaystyle
\times Z_\mu(s,\alpha,\beta \mid
s_\alpha + 1,s_\beta + 1,x_\gamma ,x_\mu)
%\end{array} \end{equation}
%$$
%\begin{array}{l} \displaystyle
\\ [1.1em]  \displaystyle
+ \phi(x_\alpha - s_\alpha) \phi(x_\beta -s_\beta)
\\ [1.1em]  \displaystyle  \times
{\rm Ad} (\tilde{v}_{s+\hat{\alpha},\alpha}(s_\alpha + 1,
x_\beta, x_\gamma, x_\mu) )
\\ [1.1em]  \displaystyle\times
Z_\mu(s+\hat{\alpha},\beta \mid
s_\alpha + 1,s_\beta + 1,x_\gamma ,x_\mu)
%\end{array}
%$$
%$$
%\begin{array}{l} \displaystyle
\\ [1.1em]  \displaystyle
+ \phi(x_\alpha - s_\alpha) \phi(x_\gamma -s_\gamma)
\\ [1.1em]  \displaystyle  \times
{\rm Ad} (\tilde{v}_{s+\hat{\alpha},\alpha}(s_\alpha + 1,
x_\beta, x_\gamma, x_\mu) )
\\ [1.1em]  \displaystyle \times
Z_\mu(s+\hat{\alpha},\gamma \mid
s_\alpha + 1,x_\beta ,s_\gamma + 1,x_\mu)
%\end{array}
%$$
%$$
%\begin{array}{l} \displaystyle
\\ [1.1em]  \displaystyle
+ \phi(x_\alpha - s_\alpha)
Z_\mu(s,\alpha \mid
s_\alpha + 1,x_\beta,x_\gamma ,x_\mu) ]  \,,
%\end{array}
%$$
\end{array} \end{equation}
where
\begin{equation}
\begin{array}{l} \displaystyle
Z_\mu(s,\alpha,\beta,\gamma \mid x)  \\  [1.1em] \displaystyle
= (\tilde{v}_{s+\hat{\gamma},\gamma} \tilde{v}_{s+\hat{\gamma}+
\hat{\beta},\beta} \tilde{v}_{s+\hat{\gamma}+\hat{\beta}+\hat{\alpha},
\alpha}  \\ [1.1em]  \displaystyle
\times \partial_\mu \tilde{v}^{-1}_{s+\hat{\gamma}+\hat{\beta}
+\hat{\alpha},\alpha} \tilde{v}^{-1}_{s+\hat{\gamma}+\hat{\beta},\beta}
\tilde{v}^{-1}_{s+\hat{\gamma},\gamma}) (x), \\ [1.1em]  \displaystyle
%\end{array}
%$$
%\begin{equation} \begin{array}{l} \displaystyle
Z_\mu(s,\alpha,\beta \mid x)  =
(\tilde{v}_{s+\hat{\alpha},\alpha} \tilde{v}_{s+\hat{\alpha}+
\hat{\beta},\beta} \\ [1.1em]  \displaystyle
\times \partial_\mu \tilde{v}^{-1}_{s+\hat{\alpha}+\hat{\beta},
\beta} \tilde{v}^{-1}_{s+\hat{\alpha},\alpha}) (x), \\ [1.1em]
%\end{array} \end{equation}
%$$
%\begin{array}{l} \displaystyle
\displaystyle
Z_\mu(s,\alpha \mid x)  =  (\tilde{v}_{s+\hat{\alpha},\alpha}
\partial_\mu \tilde{v}^{-1}_{s+\hat{\alpha},\alpha}) (x),\\ [1.1em]
{\rm Ad}(\tilde{v}) M  =  \tilde{v} M \tilde{v}^{-1}.
%\end{array}
\end{array} \end{equation}
%$$
Note that $Z_{\mu}(s,\alpha,\beta,\gamma \mid x)$ and
$Z_{\mu}(s,\alpha,\beta \mid x)$ are symmetric in
$\alpha, \beta, \gamma$ and $\alpha, \beta$, respectively, due to the
cocycle condition. It is straightforward to verify that for
$x\in \tilde{f}(s,\mu)$
\begin{equation}
{\cal P}\, {\rm exp}\{ \int_0^1 dt A_\mu^{(s)}(x+(1-t)\hat{\mu}) \}]
= u^s_{x,x+\hat{\mu}},
\end{equation}
where $u^s_{x,x+\hat{\mu}}$ are the link variables in the complete axial
gauge defined in ref. \cite{Lue}.
In eq. (\ref{eq37}) one can take the limit
$\epsilon \rightarrow 0$ and put $\phi(t) = t$ for $0 \leq t \leq 1$.

In order to obtain the global gauge field we make use of the section
$w^s$, originally defined on the boundary of the hypercube \cite{Wie}
and extended
to its interior in ref. \cite{Sch}, and write
\begin{equation} \begin{array}{l}
A_\mu(x) = w^s(x)^{-1} A_\mu^{(s)}(x) w^s(x) \\ [1.1em]
+ w^s(x)^{-1} \partial_\mu w^s(x).
\end{array} \end{equation}

Under a lattice gauge transformation,
\begin{equation}
U(s,\mu) \to
\bar{U}(s,\mu) = g(s) U(s,\mu) g(s+\hat{\mu})^{-1},
\end{equation}
the local gauge field transforms as
\begin{equation}
\bar{A}_\mu^{(s)}(x) = g(s)^{-1} A_\mu^{(s)}(x) g(s),
\end{equation}
and the section transforms as \cite{Sch}
\begin{equation}
\bar{w}^s(x) = g(s) w^s(x) g(x)^{-1},
\end{equation}
where $g(x)$ is the lattice gauge transformation extrapolated to the
interior of all the hypercubes. The latter is given in ref. \cite{Sch}.
This gives
the desired result
\begin{equation}
\bar{A}_\mu(x) = g(x) (A_\mu(x) + \partial_\mu) g(x)^{-1}.
\end{equation}

\section{AN EXAMPLE: THE CHIRAL SCHWINGER MODEL}

We shall now apply our method to the chiral
Schwinger model.
A lot is known about this model from other studies \cite{Chi}. Our
approach differs from most of the other approaches by working in
Euclidean space. Questions regarding the positivity of the transfer
matrix and the like will and can be ignored here.

The action of this model is
\begin{equation}
\int d^2x \bar{\psi} i (\p + \A P_{-}) \psi.
\end{equation}
Our choice of the $\gamma$ matrices is
\begin{equation}
\gamma_1 = \left(\begin{array}{cc}
                    & 1 \\
                  1 &
                \end{array}\right),
\gamma_2 = \left(\begin{array}{cc}
                    &-i \\
                  i &
                \end{array}\right),
\end{equation}
with
\begin{equation}
\gamma_5 = i\gamma_1\gamma_2 = \left(\begin{array}{cc}
                                     -1 &   \\
                                        & 1
                                     \end{array}\right).
\end{equation}
The gauge fields will be written $A_\mu = i A_{\mu}^0$, where
$A_{\mu}^0$ is real. This then leads to the Dirac operator
\begin{eqnarray}
i \DD (A) &=&
 i \left(\begin{array}{cc}
                    & \partial_1 + A_2^0 \\
 \partial_1 - A_2^0 &
                \end{array}\right) \nonumber \\ [1.1em]
 &+& \left(\begin{array}{cc}
                    & \partial_2 - A_1^0 \\
-\partial_2 - A_1^0 &
                \end{array}\right).
\end{eqnarray}
We may assume that $i \DD(A)$ has no zero modes due to the vanishing
of ${\rm det}\; i \DD(A)$ in this case.
Being a two-dimensional model,
it is easy to find a closed form for the
inverse Dirac operator in eqs. (\ref{eq29},\ref{eq210}), which
is the solution of
\begin{equation}
i\DD (A) (i\DD (A))^{-1} (x,y) = \delta^{2} (x-y)\; {\bf 1}.
\end{equation}
We obtain
\begin{equation} \begin{array}{l} \displaystyle
(i \DD (A))^{-1} (x,y) = - \int \frac{d^{2}k}{(2\pi)^2}
\frac{{\rm e}^{ik(x-y)}}{k^2} \\ [1.1em]  \times
%\left(\begin{array}{cc}
%                   & -k_1 + i k_2  \\
%       -k_1 -i k_2 &
%               \end{array}\right) \\ \times
\not\!{k}  \left(\begin{array}{cc}
   f^{\ast -1}(x,y) &            \\
                    & f(x,y)
                \end{array}\right),
\end{array} \end{equation}
where
\begin{equation}  \label{eq56} \begin{array}{l}
f(x,y) = {\rm exp}\{ \int d^{2}z (s(x,z)-s(y,z)) \\ [1.1em]  \times
(A_1^0 (z) +i A_2^0 (z)) \}
\end{array} \end{equation}
with
\begin{equation}
s(x,z) = \int \frac{d^{2}q}{(2\pi)^2} \frac{{\rm e}^{iq(x-z)}}{q^2}
(-q_1 + i q_2).
\end{equation}
This result generalizes to other gauge groups as well.

We are primarily interested in the gauge invariant sector of the model,
as it arises e.g. after the anomalies have been cancelled.
The (non-trivial) dynamics in this sector is driven by the
$\eta$-invariant. To keep the calculations in this write-up
short, we shall
fix the gauge and state the general result only at the end.
We choose Landau gauge, $\partial \cdot A^{0}(x)=0$.
In this case eq. (\ref{eq56}) simplifies to
\begin{equation} \begin{array}{l} \displaystyle
f(x,y) = {\rm exp}\{ i\int \frac{d^{2}q}{(2\pi)^2}
\frac{({\rm e}^{iqx}-{\rm e}^{iqy})}{q^2}\\ [1.1em]  \times
(-q_1 A_2^{0}(q) + q_2 A_1^{0}(q))\}.
\end{array} \end{equation}
On the torus (or any other compact space) the momenta are
discrete. So, whenever we write an integral over internal momenta, the
appropriate sum respecting the boundary conditions should actually be
understood. In our case it is, however, simpler to do the integrals
rather than the corresponding sums. In the infinite volume limit both
methods should give the same result.

We are now ready to compute the effective action. We choose the
point-splitting method to regularize the integrals, which is
computationally simple. Using the Pauli-Villars regularization
gives the same result. It would have been even more appropriate to
use the lattice regularization as proposed in sec. 3, in particular as
the continuum gauge fields can be stated
explicitly in this case \cite{Sch}.
However, it is hopeless to find a closed form for the inverse of the
lattice Dirac operator, which means that one has to do the sums
numerically. We shall return to  that - as a check of the method -
when we will apply our ideas to four-dimensional models.
But we believe
that the results will be the same. Writing now $^{t}\!A = i t A^0$,
we have
\begin{equation} \label{eq59} \begin{array}{l} \displaystyle
{\rm Im} S_{-}(A)
=  \lim_{\epsilon_1,\epsilon_2 \rightarrow 0}
\frac{1}{2\pi^2} \int_0^1 dt \int d^2x
 \\ [1.1em]  \displaystyle
\times \int d^2y\; \delta_{\epsilon_1,%
\epsilon_2}(x,y) \;
{\rm exp}\{ i \int_y^x ds_\mu t A_\mu^0 \} \\ [1.1em]  \displaystyle
\times   {\rm Tr}\; i\gamma_5 \A^{0}(x) \int \frac{d^{2}k}{(2\pi)^2}
  \frac{{\rm e}^{ik(x-y)}}{k^2} \\ [1.1em]  \times
 \not\!{k} \;\left(\begin{array}{cc}
   f^{\ast -t}(x,y) &            \\
                    & f(x,y)^t
                \end{array}\right),
\end{array} \end{equation}
where
\begin{equation} \begin{array}{l}
\delta_{\epsilon_1,\epsilon_2}(x,y)  \\ \displaystyle
 = \frac{\epsilon_1}{(x_1 -y_1)^2 +%
\epsilon_1^2} \;\frac{\epsilon_2}{(x_2 -y_2)^2 +\epsilon_2^2} \; .
\end{array} \end{equation}
The integral over $k$ can be done immediately and gives
\begin{equation}
\frac{i}{2\pi} \frac{\not\!{x} -\not\!{y}}{(x-y)^2}.
\end{equation}
For the integral over $y$ only the region $y \approx x$ is relevant. We
may therefore expand the line integral and $f(x,y)^t$ about $y=x$.
This gives
\begin{equation} \begin{array}{l}
{\rm exp}\{i\int_y^x ds_{\mu} t
                              A_{\mu}^0\} = 1 + i(x_1 -y_1) t A_1^{0}(x)
\\ [1.1em]
+i(x_2 -y_2) t A_2^{0}(x) + \cdots
\end{array} \end{equation}
and
\begin{equation} \begin{array}{l}
f(x,y)^t = 1 + (x_1 -y_1) t A_2^{0}(x) \\ [1.1em]
- (x_2 -y_2) t A_1^{0}(x) %
+ \cdots .
\end{array} \end{equation}
If we insert this now into eq. (\ref{eq59}) we obtain
\begin{equation} \label{eq514} \begin{array}{l} \displaystyle
{\rm Im}S_{-}(A)
= - \frac{1}{4\pi} \int_0^1 dt t \int d^2x {\rm Tr}\; %
i \gamma_5 \\ [1.1em]  \times
\left(\begin{array}{cc}
%  A_1^0(x)^2 +A_2^0(x)^2 &            \\
%                   & A_1^0(x)^2 +A_2^0(x)^2
   A^0(x)^2  &            \\
             & A^0(x)^2
                \end{array}\right)  \\ [0.5em]
                = 0
\end{array} \end{equation}
($A^0(x)^2 = A_1^0(x)^2 + A_2^0(x)^2$).
It is straightforward to repeat the calculations in an arbitrary gauge.
In momentum space we find
\begin{equation}  \begin{array}{l} \displaystyle
{\rm Im} S_{-}(A)
= \frac{1}{2\pi} \int \frac{d^2q}{(2\pi)^2} \frac{1}{q^2}
\\ [1.1em]  \displaystyle
\times (q_1 A_2^0(q) - q_2 A_1^0(q))\;  q\cdot A^0(q),
\end{array} \end{equation}
which is the anomalous term. Altogether this
means that the gauge invariant sector of the chiral
Schwinger model is identical with the corresponding vector theory,
i.e. the massless Schwinger model. In particular it says that the
$\eta$-invariant is zero.

Along the same lines we can compute the real part of the effective
action,
\begin{equation} \label{eq515}
{\rm Re}(S_{-}(A)-S_{-}(0)) = \frac{1}{2} (S(A)-S(0)).
\end{equation}
This
amounts to replacing $\gamma_5$ by $i{\bf 1}$
in eq. (\ref{eq514}), which gives
\begin{equation} \label{eq516} \begin{array}{l} \displaystyle
S(A)-S(0)
=  - \frac{1}{2\pi} \int_0^1 dt t \int d^2x \\ [1.1em]  \times
{\rm Tr}\;
 \left(\begin{array}{cc}
%  A_1^0(x)^2 +A_2^0(x)^2 &            \\
%                   & A_1^0(x)^2 +A_2^0(x)^2
   A^0(x)^2  &            \\
             & A^0(x)^2
                \end{array}\right) \\ [1.1em]  \displaystyle
          = \frac{1}{2\pi} \int d^2x \; A^0(x)^2 .
\end{array} \end{equation}
This is the well known result, which says that the gauge field
acquires a mass \cite{Jul}
\begin{equation}
m^2 = \frac{e^2}{\pi}.
\end{equation}
Note that in our notation the pure gauge field action reads
$\frac{1}{4 e^2} \int d^2x  F_{\mu\nu}^2$. It assures us that our
calculations are correct. In the chiral Schwinger model the mass is
$m^2 = e^2/2\pi$ because of eq. (\ref{eq515}). In an arbitrary gauge
eq. (\ref{eq516}) is found to be supplemented by a pure gauge term.

\section{CONCLUSIONS}

We have formulated a method, which should allow us to answer some of
the principle questions in chiral gauge theories after years of
fruitless attempts.

We have confronted our ideas with
the chiral Schwinger model and found that
the gauge invariant sector of this model equals the vector Schwinger
model. This basically means that the chiral Schwinger model has no
non-trivial dynamics.

In four dimensions the main task is now to evaluate eq. (\ref{eq214})
(or any other equivalent expression for ${\rm Im}S_{-}(A)$) and
demonstrate that the result converges as $a \rightarrow 0$. In a model
which is anomaly-free we expect the result to be gauge invariant. If
the result for ${\rm Im}S_{-}(A)$ turns out to be small, we can then
think of numerical simulations. If, however, ${\rm Im}S_{-}(A)$ is large
and strongly fluctuating, we do not see any chance for lattice
simulations at the moment. We like to emphasize that the problem of
complex action is inherent in any other approach. Whether we can
make a statement as to whether ${\rm Im}S_{-}(A)$ is non-zero or not
remains to be seen.

Finally, we like to say that it might still be possible to find a
``geometrical''
expression for ${\rm Im}S_{-}(A)$ or the $\eta$-invariant,
and one should search for it. This would improve the situation
considerably.

\begin{acknowledge}
One of us (G. S.) would like to thank the organizers of this Workshop
for a most pleasant stay in Rome.
\end{acknowledge}

%%%%%%%%%%%%%%%%%%%%%%%%%%%%%%%%%%%%%%%%%%%%%%%%%%%%%%%%%%%%%%%%%%%%%%%%

\end{document}